# Moment analysis of highway-traffic clearance distribution

Sherif M. Abuelenin, Member, IEEE, Adel Y. Abul-Magd

**Abstract**— To help with the planning of inter-vehicular communication networks, an accurate understanding of traffic behavior and traffic phase transition is required. We calculate inter-vehicle spacings from empirical data collected in a multi-lane highway in California, USA. We calculate the correlation coefficients for spacings between vehicles in individual lanes to show that the flows are independent. We determine the first four moments for individual lanes at regular time intervals, namely the mean, variance, skewness and kurtosis. We follow the evolution of these moments as the traffic condition changes from the low-density free flow to high-density congestion. We find that the higher moments of inter-vehicle spacings have a well defined dependence on the mean value. The variance of the spacing distribution monotonously increases with the mean vehicle spacing. In contrast, our analysis suggests that the skewness and kurtosis provide one of the most sensitive probes towards the search for the critical points. We find two significant results. First, the kurtosis calculated in different time intervals for different lanes varies smoothly with the skewness. They share the same behavior with the skewness and kurtosis calculated for probability density functions that depend on a single parameter. Second, the skewness and kurtosis as functions of the mean intervehicle spacing show sharp peaks at critical densities expected for transitions between different traffic phases. The data show a considerable scatter near the peak positions, which suggests that the critical behavior may depend on other parameters in addition to the traffic density.

## I. INTRODUCTION

Having a more accurate understanding of traffic behavior is becoming more required since the introduction of vehicular ad-hoc networks (VANETs) [1]. In light traffic conditions, a large transmission range is required to assure high probability of connectivity and long enough duration of connection. On the other hand, in dense traffic situations lower transmission range may be required to reduce interference [2]. The situation is more complicated in the transition phase [3]. Therefore, it is important for communicating vehicles to distinctly identify different phases of traffic and the transition points between them. In the work presented here we show the validity of statistical moments as probes for the critical points in traffic flow. Models of highway traffic often assume a functional relationship between traffic flow and traffic density [4]. The distinction between freely flowing and congested traffic in terms of traffic density is obvious. When the density of cars is high enough the system can become congested. Kerner's detailed study of empirical traffic data has led to the development of the three-phase theory of traffic [5]. According to this theory, congested traffic subdivides into two phases: 'synchronized flow' and 'wide moving jams'. The name of the phase comes from the equilibration (or synchronization) of speed and flow rate across all lanes caused by frequent vehicle lane changes. The distinction is not as obvious between the two phases of congested traffic. Wide moving jams are characterized by stopped or slowly moving vehicles within the jammed region, which widens and moves against the vehicles' direction of travel at 15-20 km/h. In contrast, the downstream front of synchronized traffic is often located at a bottleneck and traffic flow is higher than in jammed traffic—sometimes close to the rates observed in free flow. Above a critical density of vehicles, a sudden decrease in the velocity of a lead vehicle can initiate a transition from metastable states to the jamming phase. Kerner *et al*. [5], [6] study locally measured data and use a method for forecasting of traffic objects to distinguish the three phases of traffic. Traffic cellular automaton models [7], [8], [9] are used to investigate the underlying spatiotemporal patterns and the associated traffic phases. Recent analysis [10] illustrates very well that the transition from free flow to congested traffic is discontinuous. However, the theory of traffic congestion on freeways is still highly controversial despite the availability of large empirical data sets [11]. Many researchers have paid great attention to inter-vehicle spacing distribution models in order to explore the phase structure of congested traffic [12], [13], [14], [15], [16], [17], [18]. This paper presents a model-independent analysis of the empirical spacing distributions for traffic flows using methods of mathematical statistics. We study the approach of the congested traffic to the jamming limit by tracking the variance, skewness, and (excess) kurtosis [19] of the clearance distribution. The last two respectively monitor the evolution of the asymmetry and the deviation from Gaussian behavior of the distribution of headways. Besides the convenience of replacing thousands of measured spacings with three numbers, it is argued that particularly the skewness and kurtosis contain a large part of the physics embedded in the raw data in terms of their mutual relation. From previous studies of the thermodynamics of phase transitions [20], it has been seen that a system undergoing a phase transition that passes through a critical point can exhibit enhanced fluctuations. Such estimators as skewness and kurtosis should be sensitive to large fluctuations generated as the traffic density passes through (or close to) a critical point of a phase transition. Thus at the critical point, one would expect large non-monotonic behavior in these estimators. Similar techniques have been developed to address those effects in the moments of distributions of observables in various branches of physics, astronomy, biology and economy (see, e.g. [21], [22], [23], [24], [25], [26], [27], [28], [29], [30]). The paper is organized as follows. The next section is dedicated to related work. In section III we define the mean, standard deviation, skewness and kurtosis of the clearance distribution used in this study. In section IV we present the results for the moments obtained from real road measurements as a function of the mean value of the inter-vehicle spacings. A brief discussion of the limitations of the approach is also presented. Finally we summarize our study in section V.

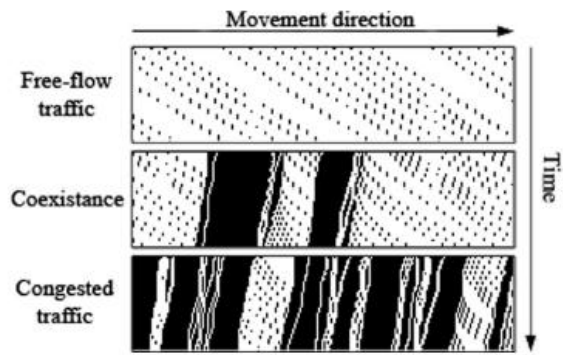

Fig. 1. Space‑time diagram showing phase transition [40].

## II. RELATED WORK

### A. Traffic modeling

The interest in understanding traffic dynamics dates back to the 1930's [4]. It became increasingly necessary to obtain mathematical description of the traffic flow process. Earlier studies [31], [32], [33] suggested that inter-vehicle headways are Poisson (exponentially) distributed [34]. The availability of large sets of empirical traffic data permitted re-evaluation of traffic flow theories and models [35]. Empirical data were used to confirm that the exponential model is generally a good fit for the distributions of inter-vehicle spacings (e.g. [36], [37]). Recent studies confirmed that this was valid for free flowing traffic [16], [38]. As for other traffic regimes, different models were proposed that provide better fits for the empirical data. This includes using Gaussian unitary ensemble (GUE) to model headway distribution in traffic jam [16]. A super-statistical approach [16], generalized-extreme-value [3] and log-normal [18] distributions were proposed to model headway distribution in transitional traffic regime. Also, shifted exponential distribution [39] was suggested to be a better model for spacing distribution in single-lane free-flow traffic. Fig. 1. [40] shows a cellular-automaton space-time diagram of the transition from free-flow to congested traffic.

### B. Kerner's traffic phases

Traditionally, highway traffic was considered to have two phases, free flow and traffic jam. Kerner [5] introduced the three-phase traffic theory to explain the empirical spatiotemporal features of traffic. According to Kerner's theory, there is a third intermediate phase, synchronous traffic. With three-phase theory, traffic breakdown cannot be related to a single classical capacity value. The infinite number of highway capacities of free flow at a bottleneck makes theory a bit harder, but it might also represent a more accurate modeling of real observations [35]. One of the consequences of the theory is that the transition between traffic regimes does not occur at specific values for density. The fundamental hypothesis of Kerner's three-phase traffic theory is that the steady states of synchronized flow cover a two-dimensional region in the flow-density plane [44]. Transitions between phases are accompanied by hysteresis effects and lead to a double Z-characteristic [5], [42]. A phase transition occurs when a critical local disturbance in an initial traffic phase appears whose amplitude exceeds some critical value. Critical amplitudes of these disturbances required for phase transitions and the dependencies of the critical disturbance amplitudes on traffic variables (e.g., on the vehicle density) differ greatly for different phase transitions. This means that these transitions exhibit probabilistic nature [42], [43]. Examples for practical reasons of such critical disturbances in free and synchronized flows could be overtaking maneuvers, fluctuations in flow rates upstream, a sudden vehicle braking, deceleration of vehicle, lane changing, any random behavior of drivers can cause this fluctuation. [43], [44].

C. Effect on inter-vehicular communications

In a VANET, vehicles act as nodes in a network. Equipped with transmitters and receivers, they exchange information in an ad-hoc manner. Information is exchanged by vehicle-tovehicle and vehicle-to-roadside communication. To guarantee the establishment of the network in a road segment, any two successive vehicles have to fall within the communication range of each other. The connectivity probability is a function of the communication range, and the headway distribution. The inter-vehicle spacing distribution is different for different traffic regimes. The network behaves differently in different traffic phases. Therefore, a single transmission mode may not perform stably under the changing VANET environment [41]. The system performance of VANETs can be improved by utilizing traffic conditions [35]. In free flow, the vehicles are sparse and drivers can freely choose their speeds; thus, they may disconnect from each other frequently. In congested flow, every vehicle will have well-connected neighborhood, but this may lead to higher interference and increased chance of data packets collisions [41], [45]. In mixed-mode traffic (transitions between phases) the situation is a bit more complicated. Also, the communication channel properties significantly vary according to traffic conditions. Different models are used to describe the channel in different traffic regimes [46], [47]. Different solutions were introduced to improve network performance by adapting to different traffic conditions (e.g. [41], [45], [48], [49], [50], [51]). Identification of traffic conditions is required for applying adaptive mechanisms. To the best of our knowledge, phase detection in current ITS systems relies on the measured value of traffic density (e.g. [40], [52]). As stated earlier, in real situations, random fluctuations play a role for the transition. Therefore, there can exist a range of values of traffic density (metastable region) for which traffic can be in different phases [5]. We need more accurate probes for detection of phases transitions.

III. OBSERVABLES

Descriptive statistics provide essential tools to quantify the main features of large data volumes in a few parameters. Central moments characterize the shape of the distribution of measurements around the mean value. The familiar moments of variance, skewness and kurtosis give indications on the dispersion, asymmetry and peakedness or weight of the tails of the distribution, respectively.

A. Notation and basic definitions

We consider the monitored spacing between successive vehicles $X$ to be a random variable, which follows a probability distribution function $P(X)$. Then the mean of measured spacing distribution µ can be related to the mean of the distribution

$$\mu = \langle X \rangle \qquad (1)$$

In practice, we deal with a finite set of measured spacings $\{X_i\}_{i=1,...,N}$ and assume that these are independent and identically distributed. The mean values are usually estimated by the sample averages

$$\langle X \rangle = \frac{1}{N}\sum_{i=1}^{N} X_i \qquad (2)$$

The standard deviation $\sigma$ of $X$ is the square root of the variance,

$$\sigma^2 = \langle (X - \mu)^2 \rangle \qquad (3)$$

It shows how much variation or dispersion from the average exists. A low standard deviation indicates that the data points tend to be very close to the mean (also called expected value); a high standard deviation indicates that the data points are spread out over a large range of values. The normalized central moments mr expressed as

$$m_r = \langle (X - \mu)^r \rangle / \sigma^r \qquad (4)$$

The skewness characterizes the asymmetry of the probability distribution function with respect to its mean value by measuring the extent to which the distribution "leans" to one side of the mean. If the bulk of the data are at the left and the right tail is stretched out, then the distribution is skewed right or positively skewed; if the peak is toward the right and the left tail is more pronounced, then the distribution is skewed left or negatively skewed. If the distribution is symmetric as in the case of a normal (Gaussian) distribution, then the distribution will have zero skewness. The kurtosis is a descriptor of the height and sharpness of the distribution peak. Higher values of kurtosis indicate a higher, sharper peak; lower values indicate a lower, less distinct peak. It is common practice to use kurtosis as a measure of whether the data are peaked or flat relative to a normal distribution. In terms of the normalized central moments, the skewness and kurtosis are respectively defined by

$$S = m_3 = \langle ((X - \mu)/\sigma)^3 \rangle \qquad (5)$$

$$\kappa = m_4 - 3 = \langle ((X - \mu)/\sigma)^4 \rangle - 3 \qquad (6)$$

The "minus 3" at the end of the last formula is often explained as a correction to make the kurtosis of the normal distribution equal to zero. Considering that $S$ and $\kappa$ vanish for a Gaussian distribution, the two statistical estimators measure large-spacing deviation from Gaussianity. Significant skewness and kurtosis clearly indicate that data are not normally distributed. It is well known that the sample mean is sensitive to outliers. Since the conventional measures of skewness and kurtosis are essentially based on sample averages, they are also sensitive to outliers. Moreover, the impact of outliers is greatly amplified in the conventional measures of skewness and kurtosis due to the fact that they are raised to the third and fourth powers. It is difficult to give a sensible interpretation to large values of these measures simply because we do not know whether the true values are indeed large or there exist some outliers. One seemingly simple solution is to eliminate the outliers from the data. In the present paper, we calculate the moments of the distribution ignoring spacings $X > \mu + 3\sigma$. It is very uncommon that statistical probability distributions accurately describe reality so far out in the tails.

B. Considerations on the relationship between skewness and kurtosis

Our main goal in this section is to put emphasis on the role of traffic density by examining the relation between skewness and kurtosis. Besides the convenience of replacing the whole traffic data made of thousands of figures with a single couple of numbers, it has been argued that $S$ and $\kappa$ do actually contain a large part of the physics embedded in the raw data in terms of their mutual relation, pointing to the fact that $S$ and $\kappa$ are roughly related by a quadratic relation

$$\kappa = A + B \cdot S + C \cdot S^2. \qquad (7)$$

Sattin *et al*. [53] show that a well-defined correlation exists between skewness $S$ and kurtosis $\kappa$ if the process under investigation depends only on a single parameter a. In this case $S = S(a)$ and $\kappa = \kappa(a)$. These authors suppose that there exists a neighborhood around $S = 0$ where the relation $S(a)$ is smooth enough to express a as a function of $S$: $a = a(S)$. Hence, $\kappa = \kappa(a(S)) \to \kappa(S)$. We may Taylor expand $\kappa$ around $S = 0$: and truncate the expansion beyond the second order term, we obtain the relation (7), where

$$A = \kappa(0), \ B = \kappa'(0), \ and \ C = \frac{1}{2}\kappa''(0). \quad (8)$$

Eq. (8) may only be valid as long as higher order terms remain negligible. Let us now move to the case where two parameters drive $S$ and $\kappa$, so that $S = S(a_1, a_2)$, and $\kappa = \kappa(a_1, a_2)$. No such inversion as in the one-parameter case is possible now. However, we can still reduce to that case when the dependence from one of the parameters (say, a1) is much fainter that from the other. Hence, expressing $a_2 = a_2(S, a_1)$ and $\kappa = \kappa(a_1, a_2(S, a_1)) \to \kappa(S, a_1)$ leads to an expression formally identical to (8):

$$\kappa(S) = \kappa(0, a_1) + \kappa'(0, a_1)S + \frac{1}{2}\kappa''(0, a_1)S^2 + \cdots \quad (9)$$

Therefore, for each fixed value of a1, the curve $\kappa(S)$ is approximately parabolic. Varying $a_1$, we plot on the plane $(S, \kappa)$ different parabolas. If the dependence from a1 is weak, we recover a fan of parabolas close to each other. Therefore, according to Sattin *et al*. [53], the dependence of congested traffic on parameters other than the traffic density can be estimated by measuring the spread of the experimental points of $\kappa(S)$ around the fitting parabola (7). In order to provide a physical rationale to the relation between skewness and kurtosis, one can be content with just a phenomenological approach, i.e., postulating an analytical approximation to the empirical probability distribution function. The data analyzed below in terms of the moments $S$, $\kappa$ have $S > 1$.

C. Statistical error estimation

As the moment analysis is a statistics hungry study, the error estimation is crucial to extract physics information from the limited experimental data. Several statistical methods have been presented. We shall use a method formulated by Luo [54]. He applies the delta theorem in statistics [19] to derive the error formula for various order moments. The delta theorem approximates the distribution of a transformation of a statistic in large samples in terms the distribution of the statistic itself. His results for the variance of the sample $\sigma$, $S$ and $\kappa$ are

$$Var(\sigma) = 1/4N\ (m_4 - 1)\sigma^2, \qquad (10)$$

$$Var(S) = 1/N\ [9 - 6m_4 + m_3^2(35 + 9m_4)/4 - 3m_3 m_5 + m_6], \qquad (11)$$

$$Var(\kappa) = 1/N\ [-m_4^2 + 4m_4^3 + 16m_3^2(1 + m_4) - 8m_3 m_5 - 4m_4 m_6 + m_8], \qquad (12)$$

where the normalized central moments $m_r = \langle (X - \mu)^r \rangle / \sigma^r$. In terms of the normalized central moments $S = m_3$ and $\kappa = m_4 - 3$.

## IV. DATA ANALYSIS

### A. Data sets

The empirical vehicles data used in this paper are obtained from Berkeley Highway Laboratory (more detailed reports are found in [55]). The data are collected using dual loop detectors installed on the five-lane interstate I-80 road in California. There are 8 dual-loop stations distributed along a 2.7 mile road segment. Each station is a pair of inductive loop detectors, one upstream and one downstream in the same lane. Each sensor is effectively a square. They are six feet across, and their centers are twenty feet apart. Each vehicle stream data file record indicates a matched pair of upstream and downstream transitions from a specific lane at a specific station. Stations process real-time individual vehicle detection data. The arrival instants are accurate within 1/60 of a second. Wisitpongphan *et al.* [56] calculated inter-arrival time and speed statistics at three different time periods on Highway I-80, taken from the Berkeley data. They proposed a linear relationship between the road-level inter-vehicle spacing and the product of road-level inter-arrival time between vehicles and vehicles' speed. In the present work, we calculate the spacing between successive vehicles passing in the same lane. This is performed as follows. First, we find the difference between the upstream and downstream on-times (or equivalently off-times), which represents the travelling time of the vehicle between the two sensors of the station. Knowing the distance separating the two sensors, we determine the velocity of each vehicle by dividing the distance over the time. Next, we compute the spacing between each two consecutive vehicles (the rear bumper-to-rear bumper distance) through multiplying each vehicle's velocity and the time headway (the time difference between the crossing time of the vehicle to one of the sensors, and the next time instant the sensor is on). Road-level inter-vehicle spacing X can be approximated as the product of road-level inter-arrival time between vehicles and vehicles' speed, as suggested in [56]. In this article we use the data collected using one station on the 26th of June, 2006 [55]. Eleven different time periods with different traffic volumes are carefully selected for our case studies, with each time period lasting for one hour: i) Morning traffic with low traffic volume and high speed (4 a.m. - 7 a.m.) ; ii) Non-rush hour traffic with moderate traffic volume and high speed (10 a.m. - 12 p.m.); iii) Rush-hour traffic with low speed, some intermittent congestion, and with high traffic volume (7 a.m. - 10 a.m. and 4 p.m. - 7 p.m.).

B. Correlation of traffic densities in neighboring lanes

In this work, we consider the single-lane traffic behavior by extracting traffic information in a single lane from the 5-lane data [55], assuming that the traffic flows in neighboring lanes are not correlated. To examine this assumption, we have calculated the correlation coefficients for inter-vehicle spacings $X^\alpha$, $X^\beta$ in different lanes $\alpha, \beta$

$$r_{\alpha,\beta} = \frac{\sum_{i=1}^{N_{\alpha,\beta}}(X_i^\alpha - \mu^\alpha)\left(X_i^\beta - \mu^\beta\right)}{\sqrt{\sum_{i=1}^{N_{\alpha,\beta}}(X_i^\alpha - \mu^\alpha)^2 \sum_{i=1}^{N_{\alpha,\beta}}(X_i^\beta - \mu^\beta)^2}} \quad (13)$$

in three one-hour periods starting at 5 a.m., 8 a.m. and 5 p.m., where $N_{\alpha,\beta} = \min(N^\alpha, N^\beta)$ and $N^\gamma$ is the number of spacings between vehicles in lane $\gamma$. The mean spacings in all lanes in each of these time intervals are 412.1 m, 137.7 m, and 37.7 m. The obtained values of the correlation coefficients are given in Table I. The mean values of the correlation coefficients are $0.007 \pm 0.057$, $-0.008 \pm 0.32$ and $0.021 \pm 0.030$, respectively, where the errors are given by the corresponding standard deviations. We have no reason to expect more correlations in other time intervals. Thus, the independence of traffic in neighboring lanes is a reasonable assumption. This also indicates that traffic density is the fundamental variable that determines not only the traffic's average properties, such as the mean space clearance $\mu$, but also fluctuations around those averages.

Table I. Correlation coefficients for inter-vehicle spacings in different lanes α, β

| α, β | 5 a.m | 8 a.m | 5 p.m |
|---|---|---|---|
| L0,L1 | -0.093 | -0.067 | -0.023 |
| L0,L2 | 0.068 | -0.043 | 0.024 |
| L0,L3 | -0.076 | -0.048 | 0.019 |
| L0,L4 | -0.024 | 0.038 | 0.002 |
| L1,L2 | 0.007 | -0.002 | 0.012 |
| L1,L3 | 0.005 | -0.001 | -0.002 |
| L1,L4 | 0.102 | 0.020 | -0.010 |
| L2,L3 | -0.001 | 0.003 | 0.056 |
| L2,L4 | 0.038 | 0.017 | 0.072 |
| L3,L4 | 0.04 | 0.004 | 0.054 |

C. Variance, skewness and kurtosis of the 5-lane traffic data

We then consider the density dependence of higher moments of inter-vehicle spacing distributions in individual lanes. We compute the measures of variance, skewness and kurtosis using all spacings less than $\mu + 3\sigma$. The main result of this section is summarized in Fig. 2-5 below. Fig. 2 shows the relationship between the skewness and kurtosis. Solid curves are interpolating second-degree polynomials (7), and the values of the coefficients are explicitly displayed. The data do indeed distribute around the quadratic curve (7). As follows from Eq. (9), the quadratic relationship between $S$ and $\kappa$ suggests that both of these statistics are sensitive mainly to the variation of a single parameter, which may be related to the traffic density. The three moments $\sigma$, $S$, and $\kappa$ are plotted against the mean spacing $\mu$ in Fig. 3 and 4. Several conclusions may be drawn from the plots: data belonging to different lanes at different time intervals do indeed distribute around smooth curves. Loosely speaking, results may be parametrized according to the degree of order that is commonly associated to traffic density. The relationship between $\sigma$ and $\mu$ is monotonic and almost linear. This means that the traditional measure of spacing distributions, variance (or standard deviation), has failed to capture the presence of phase transition(s). In contrast, the evolution of $S$ and $\kappa$ with increasing traffic density is by no means monotonic. They both have rather complex phenomenological density dependence patterns. Each of the two moments has a sharp peak that spreads over a finite range in the vicinity of $\mu_c = 45 \pm 3$ m ($\rho \sim 22.2$ vehicle/km). This reveals the existence of a critical mean spacing clearance $\mu_c$, as a phase transition that passes through a critical point can exhibit enhanced fluctuations. Above $\mu_c$, the flow of vehicles is less impeded. There is a less evidence for another peak defined at $\mu_c` \sim 75$ m ($\rho \sim 13.3$ vehicle/km), which may indicate another phase transition. The critical spacings $\mu_c$ and $\mu_c`$ may indicate clearance for the transition between the free flow phase, the synchronized flow and the wide moving jam in Kerner's three phase theory. The fat tails indicates a high probability of large inter-vehicle spacings, this is mainly due to the values belonging to the HOV lane. Fig. 5 shows the relationship between the traffic density and the skewness and kurtosis of the spacing distributions. We note that the scatter of the experimental points in the vicinity of the position of peak(s) of $S$ and $\kappa$ which suggests that the corresponding phase transition(s) depend on other properties of the traffic in addition to its density. As discussed earlier, this includes fluctuations due to the probabilistic nature associated with phase transitions. Fig. 6. shows $S$, and $\kappa$ plotted against the density, for data taken from a different day (23$^{rd}$ of June, 2006). The figure confirms the existence of sharp peaks at density values close to those observed with the first data set.

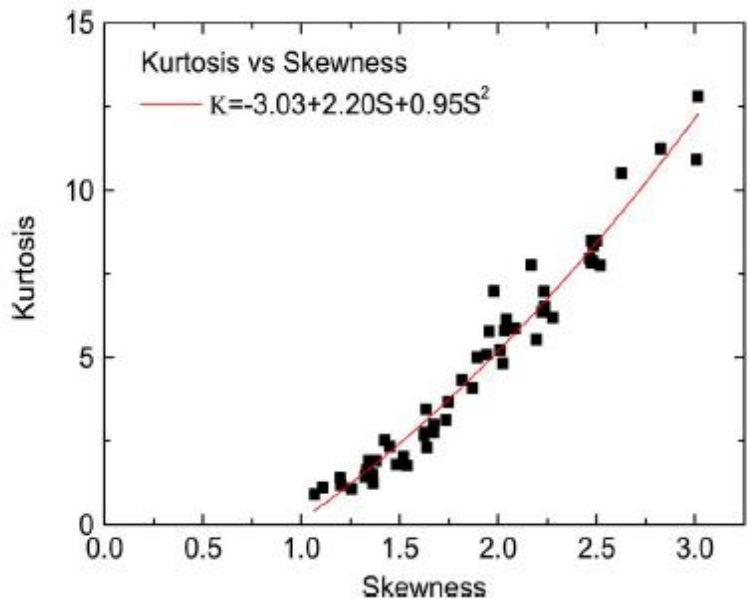

Fig. 2. The kurtosis of the traffic clearance distribution as a function of the skewness for individual lanes. The solid lines represent the second-order polynomial fits.

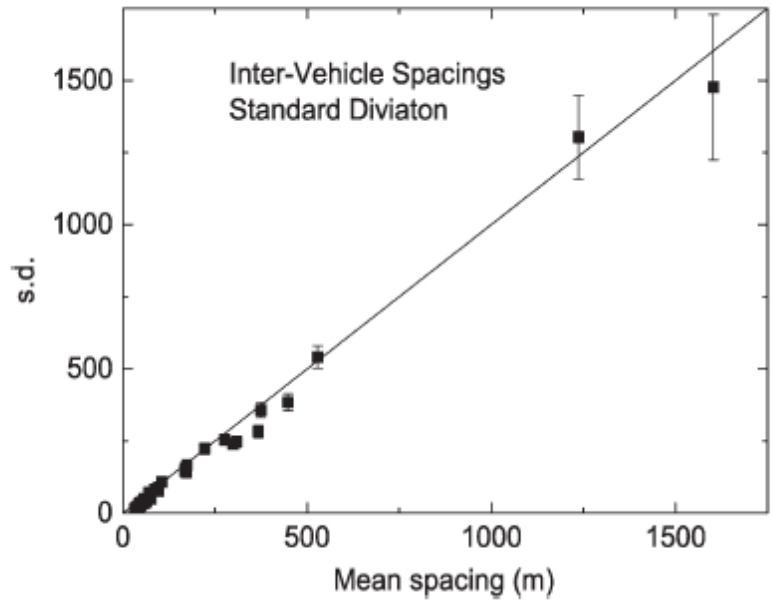

Fig. 3. The standard deviation of the traffic clearance distribution as a function of its mean value.

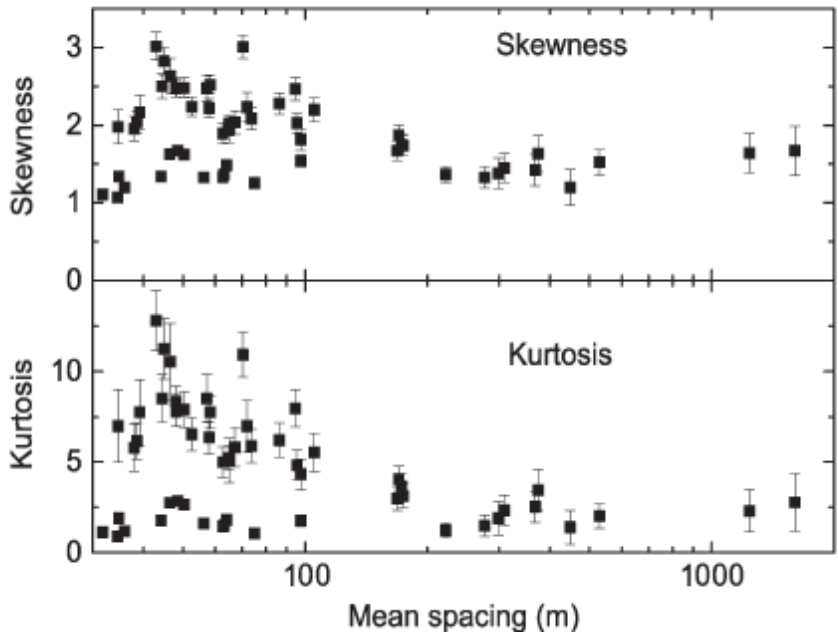

Fig. 4. The skewness and kurtosis of the traffic clearance distribution as a function of its mean value.

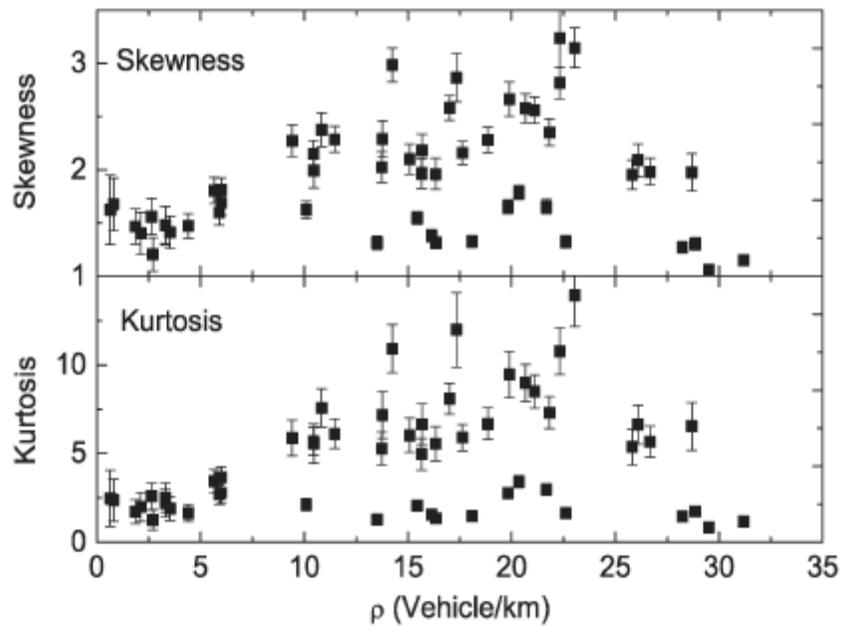

Fig. 5. The skewness and kurtosis of the traffic clearance distribution as a function of the traffic density.

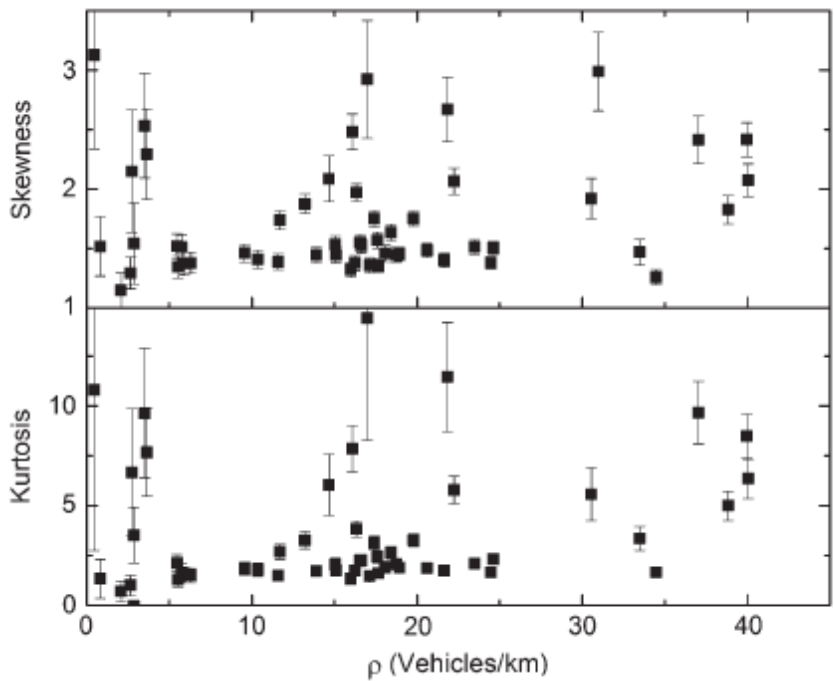

Fig. 6. The skewness and kurtosis of the traffic clearance distribution as a function of the traffic density, for a different data set (23rd of June, 2006).

V. CONCLUSION

We investigate the fluctuations of vehicular traffic density in single-lane traffic directly derived from empirical data collected in a 5-lane highway in USA. We use empirical data on inter-arrival times to calculate inter-vehicle spacings between vehicles on each lane of the road in 10 one-hour intervals. We assume that traffic flows in neighboring lanes are independent. This assumption is supported by calculating the correlation coefficients of inter-vehicle spacings in different lanes. We therefore consider that the moments of the spacing distribution in individual lanes mainly depend on a single parameter, which may be related to the mean vehicle spacing within the lane. We also show that the empirical values of the kurtosis, calculated in different time intervals for different lanes of the highway, follow a smooth function of the skewness, as required by spacing probability distributions that depend mainly on a single parameter. We then focus on the variation of the standard deviation, skewness and kurtosis with the traffic density. We find that the standard deviation of the spacing distribution increases monotonically with increasing mean spacing. In contrast the skewness and kurtosis show peaks which are "evanescent" in the sense that they will disappear in the large density jamming limit. This result validates the assumption that the use of measures of skewness and kurtosis reveals interesting evidence of true critical density dependencies. Thus, they should be good observables providing sufficient information about the critical density at the onset of traffic phase transitions. We finally note that the skewness of all of the distributions under considerations are greater than 1. The fat tails indicate a high probability of large inter-vehicle spacings due to the data points from the high occupancy vehicles (HOV) lane. These extreme events should be considered in assessing the connectivity of the networks.